# An approach for improving the distorted structured light in holographic optical tweezers

Yida Song, Zhengshu Zhang, Yi Shen, and Xionggui Tang[*]

Department of Physics, Key Laboratory of Low Dimensional Quantum Structures and Quantum Control of Ministry of Education, Hunan Normal University, Changsha, 410081, P.R. China

**Abstract:** Optical tweezers have been widely used for optical manipulation of various particles. At present, there are different type of optical tweezers. Among them, holographic optical tweezers have attracted growing attention as a powerful tools for optical trapping, optical transportation and optical sorting in many fields, due to its excellent properties including great flexibility and high convenience. Experimentally, however, the structured light has been easily distorted, which would lead to serious degradation of optical manipulation performance. In this work, the distortion of structured light is theoretically analyzed. In the following, the distortion of structured light are numerically simulated and experimentally measured. It shows that the simulated results are in consistent with the experimental ones. Then, an approach for decreasing its optical distortion is proposed, and the results reveal that the distortion of structured light can be effectively corrected. Accordingly, our study provides a way for improving the distorted structured light, which is useful for optically manipulating various particles in optical tweezers.

**Keywords**: holographic optical tweezers; structured light; optical distortion; optical manipulation.

## 1. Introduction

Optical tweezers are a powerful tool for optically manipulating microscopic particles by using intensity gradient force or phase gradient force, to realize optical trapping, optical transportation and optical sorting. Nowadays, optical tweezers have been widely used in many fields including biology, physics and soft condensed matter, due to advantages such as non-contact and high accuracy [1,2]. Since the invention of conventional optical tweezers by Ashkin A, optical manipulation technique has been developed rapidly, and various apparatuses for optically manipulating particles have been demonstrated, including fiber-based optical tweezers, plasmonic optical tweezers, waveguide-based optical tweezers and holographic optical tweezers

[*] Corresponding author. E-mail address: tangxg@hunnu.edu.cn (Xionggui Tang).

[3,4]. Among them, holographic optical tweezers have become one of the most popular platforms for optical manipulation, because of having dynamic control with high flexibility, which can create various structured light and realize versatile multifunctional optical manipulation[5,6]. In holographic optical tweezers system, a phase-only spatial light modulator (SLM) has been usually employed to purposely modulate phase profile of beams by loading the designed phase-only hologram. Frankly, the quality of hologram has great effect on the quality of structured light on the plane of optical manipulation, which strongly affects the performance of optical manipulation in holographic optical tweezers[7,8].

In the past two decades, several methods for designing holography have been proposed, which includes Gerchburg-Saxton algorithm, adaptive-additive algorithm etc [9-13]. For these algorithms, however, the phase distribution of structured light is usually poor, which would be unfavorable for manipulating particles by phase gradient force. Later, we previously proposed a method for rapidly creating structured light with high-quality phase distribution, in which a random phase factor is introduced into discrete inverse Fourier transform for calculate holograms [14,15]. In the experimental process, the generated structured light would be in good agreement with the simulated ones, if the optical system is in well adjustment. Otherwise, the quality of structured light would be obviously degraded, which directly decreases the performance of optical manipulation. Actually, the adjustment quality of optical system closely relies on personal experimental skill, which is generally time consuming. In addition, if the stability of optical platform is relatively poor, it leads to further degradation of adjustment quality of optical system. In the case, the optical system needs to be repeatedly adjusted, which has a negative effect on optically manipulating particles.

To address this issue, we propose a method for reducing the distortion of structured light by directly modifying the holograms. The simulated results and experimental ones demonstrate that the distortion of structured light can be effectively decreased. Naturally, it exhibits advantages such as simple, convenient, and time-saving. Consequently, it is very helpful for improving the performance of optically manipulating various particles.

**2. The distortion of structured light**

Up to now, different experimental setups of holographic optical tweezers have been demonstrated to realize versatile functions of optical manipulation. For our experimental setup, its schematic diagram is depicted in Fig. 1(a). A input laser beam is operated in a TEM$_{00}$ Gaussian mode. A phase-only SLM is employed to create the designed hologram, which is used for modulation incident light beams at input plane, to generate structured light at output plane where particles are manipulated. Theoretically, there is relationship of Fourier transform between optical fields at the input and output plane. Frankly, the optical components in experimental system are difficult to be well adjusted, which would lead to relatively large discrepancy between the targeted structured light and practical structured one. In the case, the distorted light beam is caused by fact that light beam arriving at the surface of SLM is non-ideal. Mathematically, the ideal light beam incident on the SLM is a plane wave, but the non-ideal light beam has special optical field distribution similar to Gaussian beam.

Firstly, how the distorted light beams appear is briefly analyzed. In the experiment, the laser beam incident on SLM is non-ideal, and then its optical field can be written as,

$$p(x,y) = E_0 \exp[-\frac{(x-x_0)^2 + (y-y_0)^2}{\omega_0^2}]\exp[jk\frac{(x-x_0)^2 + (y-y_0)^2}{2R_0^2}] \quad (1)$$

where $E_0$ denotes the maximum amplitude of light beam, and $\omega_0$ stands for its waist radius, and $R_0$ indicates curvature radius of phase profile, and $(x_0, y_0)$ is a coordinate offset between center of light beam and center of SLM, as shown in Fig. 1(b). Its radial offset can be obtained by $\Delta s = \sqrt{x_0^2 + y_0^2}$. The phase function modulated by SLM is assumed as $H(x,y)$. Accordingly, the optical field of structured light at output plane is given by using Fourier transform,

$$U(x',y') = F\{p(x,y)H(x,y)\} \quad (2)$$

Where $F\{\ \}$ denotes the operation of Fourier transform. As seen from Eq.(2), the optical field of structured light not only highly depends on phase function $H(x,y)$ of hologram, but strongly relies on the optical field $p(x,y)$ of the incident light beam. Therefore, the distortion of structured light inevitably appears. If the incident light beam is ideal, its optical field $p(x,y)$ can be considered to be equal to $E_0$. Naturally, there is a direct relationship of Fourier transform between phase function $H(x,y)$ of hologram at input plane and the optical field of structured light at output plane, which can be expressed as,

$$U(x',y') = F\{H(x,y)\} \quad (3)$$

It shows that the optical field of structured light is only related to the designed hologram.

In the following, the distortion of structured light effected by the incident light beam given in Eq. (1) are numerically investigated. Here, perfect optical vortex is chosen as an example, which has very narrow ring and fixed radius independent of their topological charges. Its topological charge is set to be equal to 25. The holograms used for creating perfect optical vortices are designed by our proposed method [14]. When intensity profile of incident light beam is only non-ideal, the simulated results are shown in Fig. 2(a-c), respectively, in which results in left side represent intensity profiles and results in right side denote phase profiles. In this case, $\omega_0 = 9.0$mm, 7.5mm, 6.0mm, respectively, but $\Delta s = 0$, and $R_0 \to \infty$ in Eq. (1). It finds that both intensity and phase profiles of structured light almost have no distortion. Next, Fig. 3(a-c) depicts the simulated results, when phase profile of incident light beam varies, i.e., $\omega_0 = 6.0 mm$, $\Delta s = 0$, but $R_0 = 60$mm, 50mm, 40mm, respectively. Similarly, it indicates that the distortion of structured light are very low. However, when $\omega_0 = 6.0 mm$, $R_0 \to \infty$, but $\Delta s = 1.0$mm, 2.0mm, 3.0mm, respectively, the simulated results demonstrate that the intensity distortion increases with the offset coordinate value grows, as given in Fig. 4(a-c), respectively.

Furthermore, the distortion of perfect optical vortices with different topological charges are discussed, when, $\omega_0 = 6.0 mm$, $R_0 = 40 mm$, but $\Delta s = 3.0$mm, respectively. The simulated results are presented in Fig. 5(a-f). These perfect optical vortices are set to have same radius, and their topological charge is 5, 15, 25, -5, -15, -25, respectively. Obviously, we finds that the intensity distortion of perfect optical vortices obviously increase, as absolute value of topological charge grows. In addition, the distortion is located in the right side of ring when its topological charge is larger than zero, but the distortion is located in the left side of ring when its topological charge is smaller than zero. Consequently, the magnitude and location of distortion are highly related to the value and direction of orbital angular momentum in perfect optical vortices. Their corresponding experimental results are provided in Fig. 6. As seen from Fig. 5 and Fig. 6, it indicates that the experimental results are in consistent with the simulated ones. In addition, it needs to point out that their corresponding phase profiles aren't given in Fig. 5, because the phase profiles almost remain unaffected.

## 3. Distortion correction of structured light

As seen from the simulated and experimental results shown above, the distortion magnitude mainly originates from not only radial offset value of incident light beam, but also orbital angular momentum of structured light. Substantially, its distortion magnitude heavily relies on frequency spectrum distribution of optical field after modulated by SLM, according to theory of information optics. Therefore, it is feasible to effectively reduce the distortion of structured light by using modifying the hologram. In the situation, flow chart for obtaining the modified hologram is presented in Fig 7. First, the initial hologram is calculated by using our proposed method [14], according to the targeted structured light $U_t(x', y')$. Then, the distorted structured light can be easily obtained by using Eq. (2). Naturally, its distortion magnitude can be obtained as follows,

$$\Delta(x', y') = U(x', y') - U_t(x', y') \qquad (4)$$

In the following calculation, the modified hologram is computed in the same way, in which the targeted structured light is replaced by a modified value,

$$U_t^i(x', y') = U_t(x', y') + k_0 + k_1 \Delta + k_2 \Delta^2 \qquad (5)$$

in which $i$ denotes the number of optimization, and ($k_0$, $k_1$, $k_2$) stands for the parameters obtained by optimization algorithm such as annealing algorithm. In calculation process, it needs to mention that the root mean square of distortion magnitude $\Delta(x', y')$ is taken as a merit function to evaluate the distortion of structured light. If its merit function is satisfied, the modified hologram is finally obtained. Otherwise, next optimization calculation for modified hologram continues.

Here, perfect optical vortices shown in Fig. 6(a-c) are selected as an example for reducing the distortion. The corresponding results are given in Fig. 8, in which Fig. 8(a-c) shows the simulated results and Fig.8(d-f) presents the experimental results. It indicates that the simulated results are in good agreement with experimental ones, which demonstrates that the quality of perfect optical vortices have been clearly improved. To further verify our method, moreover, improvement of the distortion in other structured light is carried out. The rectangle optical pattern with topological charge of 10 is selected as an example. The related results are shown in Fig. 9(a-d). Fig. 9 (a) and (c) give its simulated result and experimental one, respectively, when the hologram isn't modified. Fig. 9 (b) and (d) provide the corresponding improved intensity distribution while its hologram is modified. From Fig. 9 (a) and (c), the simulated result agrees with the corresponding experimental one. Similarly, the intensity distribution in Fig. 9 (b) is in consistent with one in Fig. 9 (d). It reveals that the distortion of rectangle optical pattern can be effectively reduced when the hologram is modified. Therefore, the proposed method is useful for improving the distorted structured light, which can promote the performance of optical manipulation in holographic optical tweezers.

## 4. Conclusions

In this work, the distortion of structured light in holographic optical tweezers have been theoretically analyzed, and numerical simulation have been carried out. Afterwards, a scheme for improving the quality of structured light has been proposed. It exhibits several advantages including simple, convenient, and time-saving. Then, the numerical and experimental results have been demonstrated, and it finds that distortion of structured light can be effectively improved. As a result, the corrected light beams are very helpful for achieving high-quality optical manipulation of small particles.

**Acknowledgments**

This work was supported by the Scientific Research Foundation of Hunan Provincial Education Department under Grant 20A315, and the Scientific Research Foundation of Changsha City under Grant kq2202238.

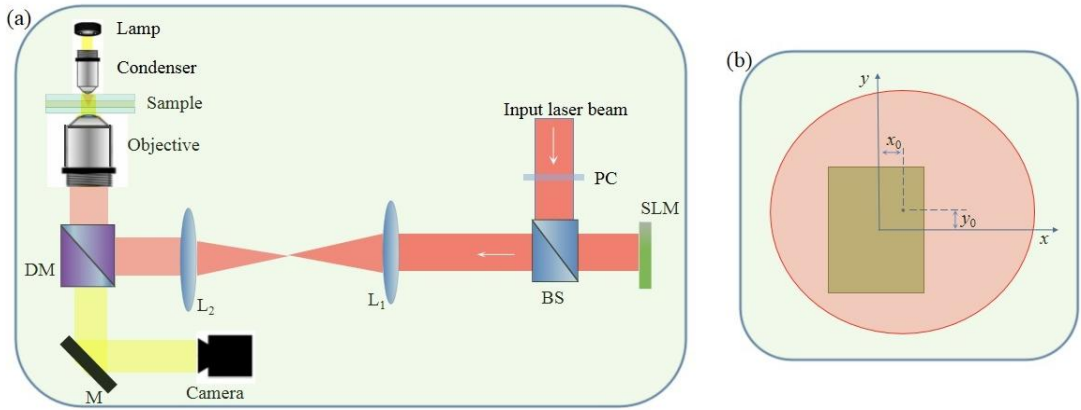

**Fig. 1** (a) schematic diagram of the experimental setup, (b) the location relationship of SLM and the incident laser beam. $L_1$-$L_2$, Lens; M, Mirror; SLM, Spatial light modulator; DM, Dichroic Mirror; BS, beam splitter; PC, Polarization controller.

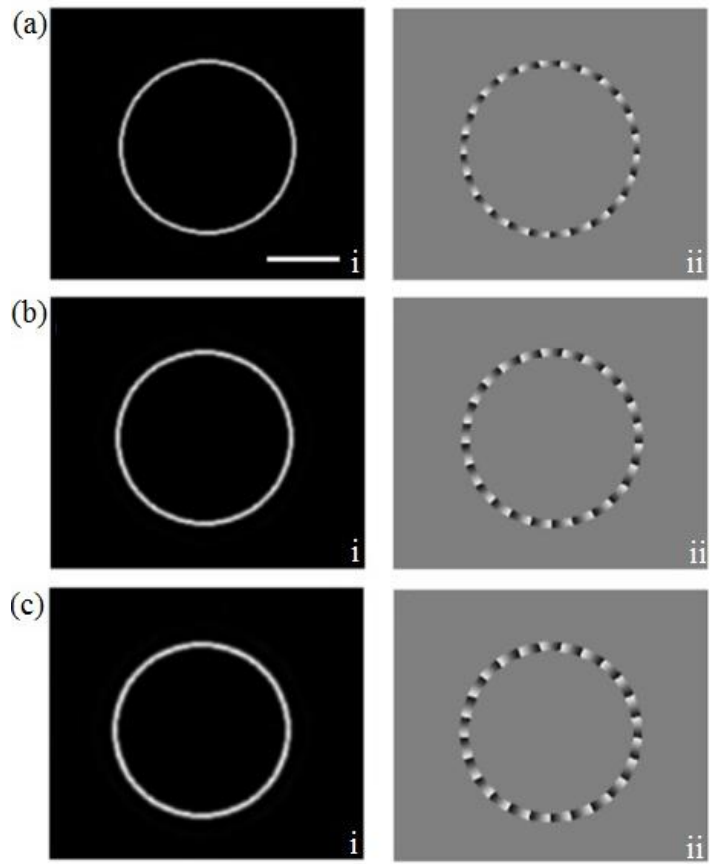

**Fig. 2** Intensity and phase profiles of perfect optical vortices as $\omega_0$ is equal to (a) 9.0mm, (b) 7.5mm ,(c) 6.0mm, respectively. The scale bar is 5 μm.

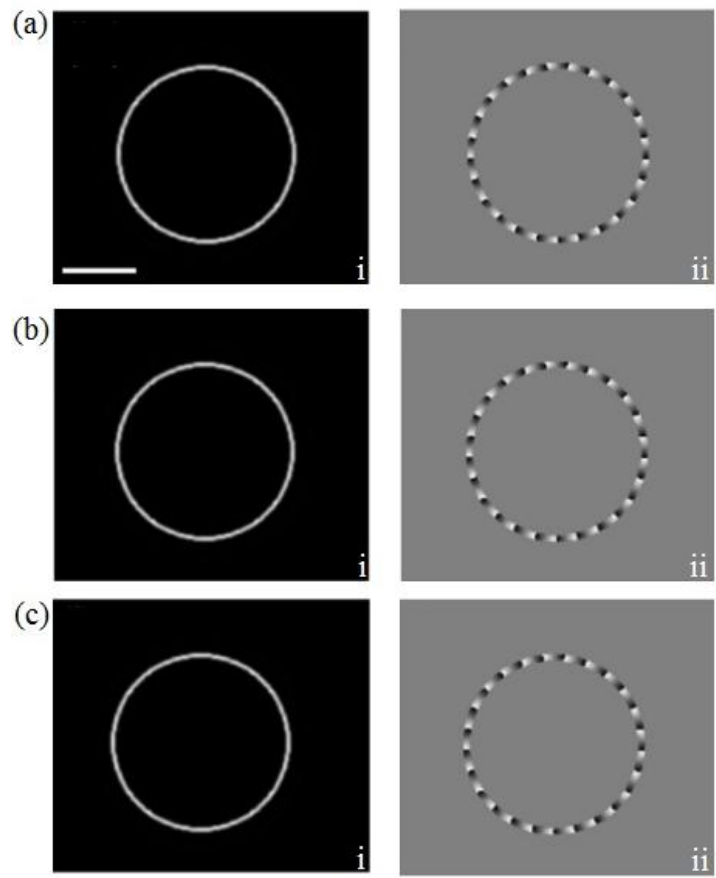

**Fig. 3** Intensity and phase profiles of perfect optical vortices as $R_0$ is equal to (a) 60mm, (b) 50mm, (c) 40mm, respectively. The scale bar is 5 μm.

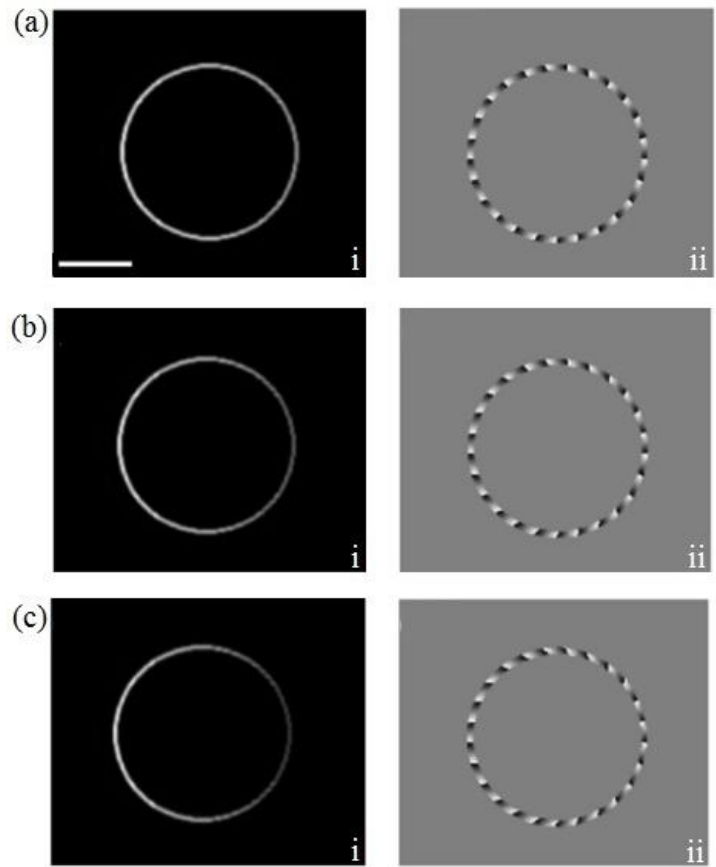

**Fig. 4** Intensity and phase profiles of perfect optical vortices as $\Delta s$ is equal to (a) 1.0mm, (b) 2.0mm, (c) 3.0mm, respectively. The scale bar is 5 μm.

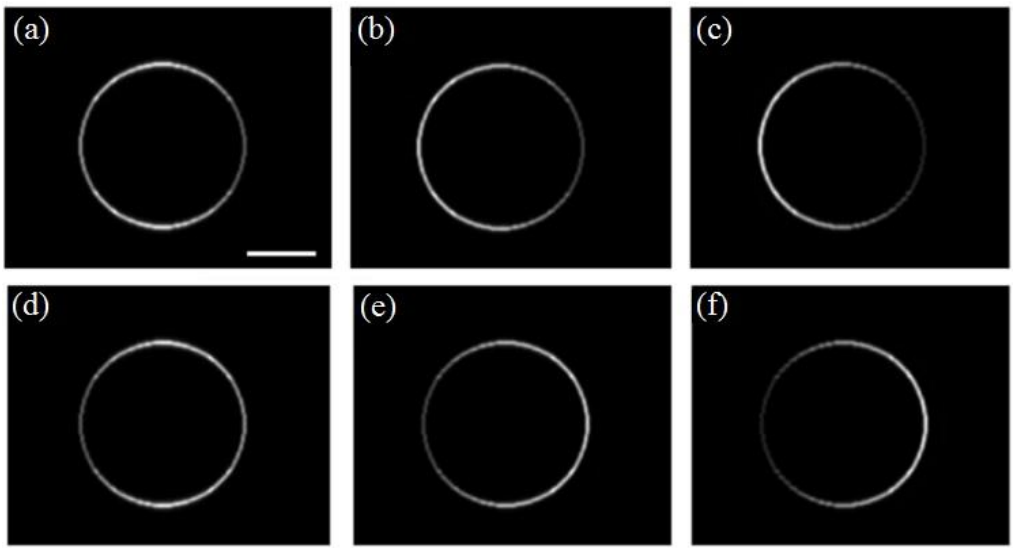

**Fig. 5** The simulated intensity distortion of perfect optical vortices with different orbital angular momentum, its topological charge is equal to (a) l=5, (b) l=15, (c) l=25, (d) l=-5, (e) l=-15, (f) l=-25, respectively. The scale bar is 5 μm.

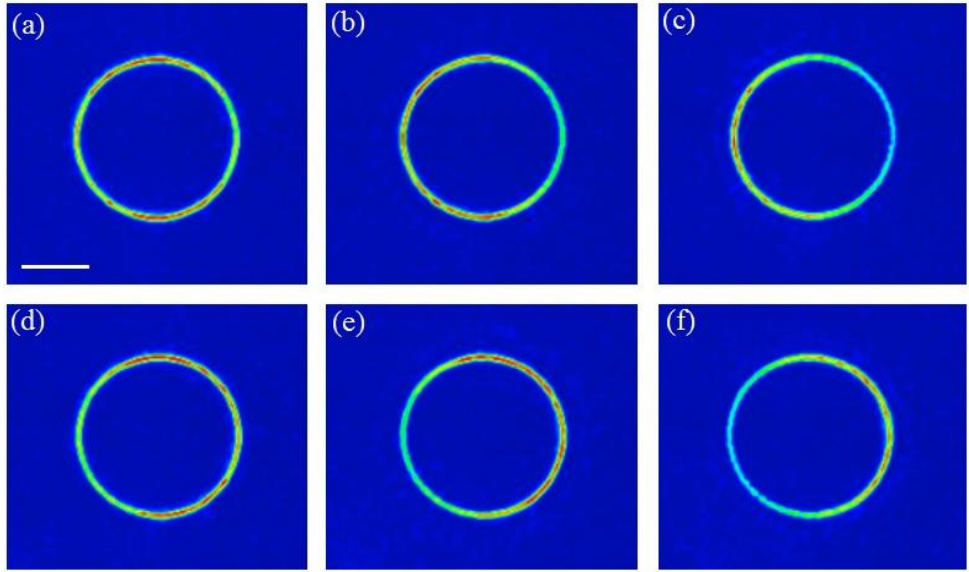

**Fig. 6** The measured intensity distortion of perfect optical vortices with different orbital angular momentum, its topological charge is equal to(a) l=5, (b) l=15, (c) l=25, (d) l=-5, (e) l=-15, (f) l=-25, respectively. The scale bar is 5 μm.

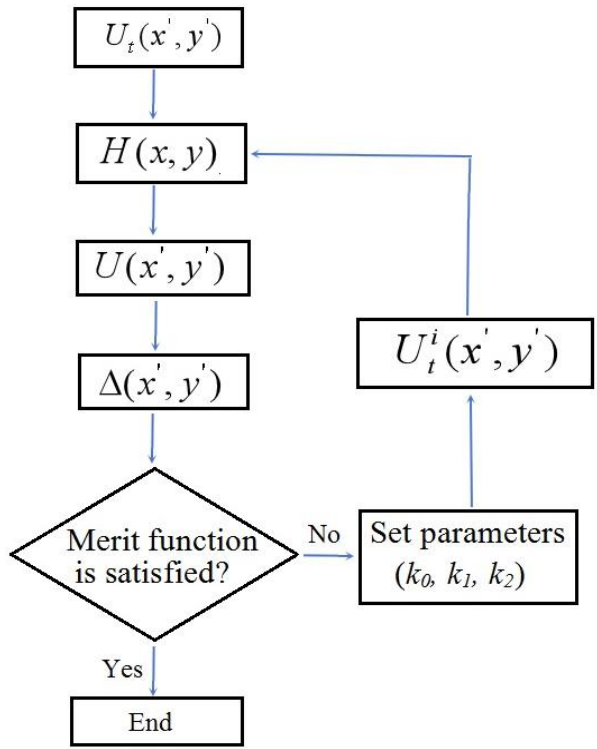

**Fig. 7** flow chart for obtaining the modified hologram.

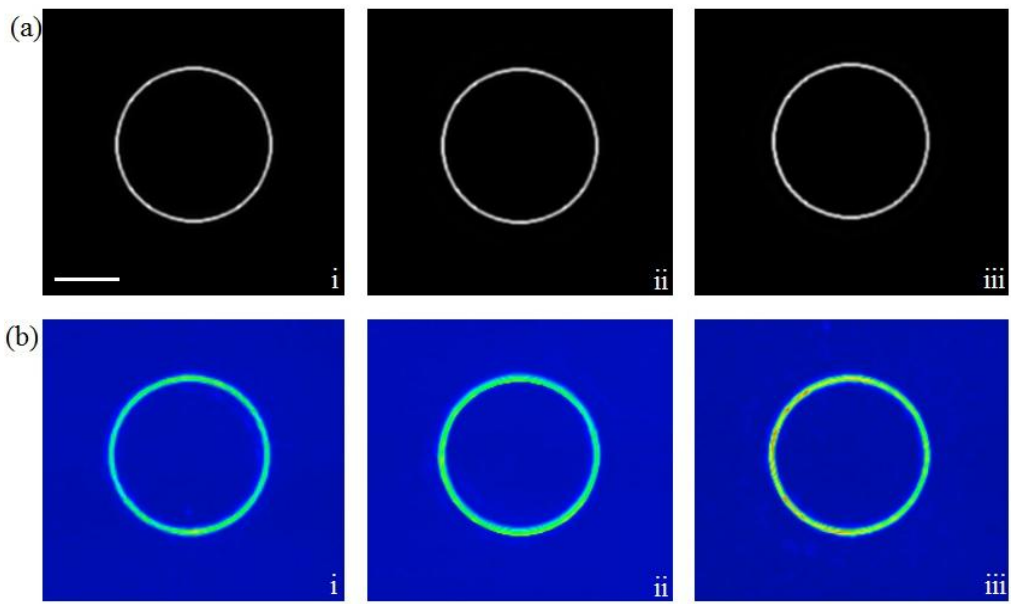

**Fig. 8** The intensity distribution after correction, (a) the simulated results and (b) the measured results. Note that panels i-iii correspond to its topological charge l=5, 15, 25, respectively. The scale bar is 5 μm.

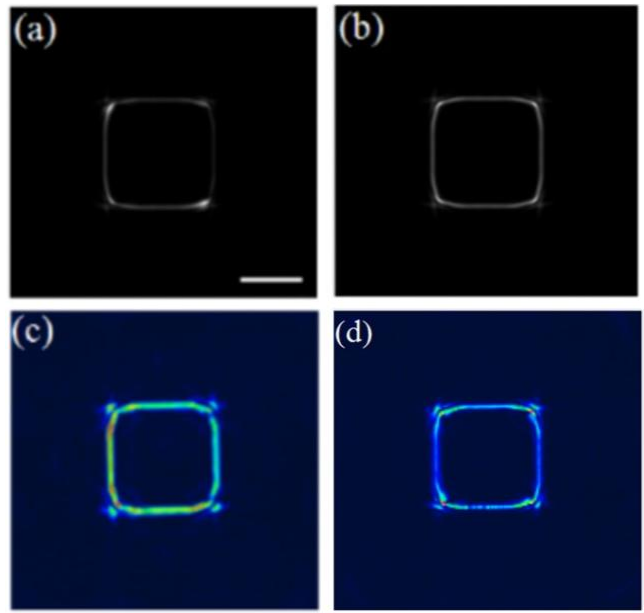

**Fig. 9** The intensity distribution of rectangle pattern, (a) the simulated result before correction; (b) the simulated result after correction; (c) the measured result before correction; (d) the measured result after correction. The scale bar is 5 μm.